\documentclass{article}
\usepackage{slashed,feynmf,epsfig,amsmath,amssymb,enumitem}
\usepackage[papersize={8.5in,11in}]{geometry}
\renewcommand{\vec}[1]{\boldsymbol{\mathrm{#1}}}
\usepackage{graphicx}
\begin{document}
\title{Scalar and Vector Field Constraints, Deflection of Light and Lensing in Modified Gravity (MOG)}
\author{J. W. Moffat\\~\\
Perimeter Institute for Theoretical Physics, Waterloo, Ontario N2L 2Y5, Canada\\
and\\
Department of Physics and Astronomy, University of Waterloo, Waterloo,\\
Ontario N2L 3G1, Canada}
\maketitle

%\date{\today}

%\thanks{PACS: 98.80.C; 04.20.G; 04.40.-b}

% ----------------------------------------------------------------

\begin{abstract}
A conformal coupling of the metric in the Jordan frame to the energy-momentum tensor, screens the scalar field gravitational coupling strength $G$ in modified gravity (MOG). The scalar field acquires a mass which depends on the local matter density: the scalar field particle is massive for the Sun and earth, where the density is high compared to low density environments in cosmology and astrophysics. Together with the screening of the vector field $\phi_\mu$, this guarantees that solar system tests of gravity are satisfied. The conformal metric is coupled to the electromagnetic matter field and energy-momentum tensor, screening $G$ for the Sun and the deflection of light by the Sun and the Shapiro time delay in MOG are in agreement with general relativity. For galaxies and galactic clusters the enhanced gravitational coupling constant $G$ leads to agreement with gravitational lensing without dark matter. For compact binary pulsars the screening of $G$ removes the monopole and dipole gravitational radiation modes in agreement with the binary pulsar timing data.

\end{abstract}

\maketitle

\section{Introduction}

The Scalar-Tensor-Vector (STVG) modified gravitational (MOG) theory~\cite{Moffat1} has successfully explained the rotation curves of galaxies~\cite{MoffatRahvar1} and the dynamics of galactic clusters~\cite{MoffatRahvar2}, as well as describing the growth of structure, the matter power spectrum and the cosmic microwave background (CMB) acoustical power spectrum~\cite{Moffat2}. The modified Newtonian acceleration law obtained in the theory in the weak field, non-relativistic approximation to the field equations, reduces to Newtonian gravity for the solar system describing solar system experiments in agreement with general relativity (GR). A conformal metric ${\tilde g}_{\mu\nu}$ is introduced with a screening mechanism dependent on the density of matter that suppresses the strength of gravitational interaction described by the scalar field $G$, for compact, dense objects such as the Sun and earth, while allowing for an increase of the gravitational interaction experienced by stars in galaxies and galactic clusters. The null geodesic equation for photons determines photon paths in the Jordan frame conformal metric, and in the Einstein frame metric the gravitational constant is screened, yielding the deflection of light by the Sun, and the Shapiro time delay in agreement with GR. The enhanced gravitational interaction experienced by photons in the lensing of galaxies and galactic clusters leads to agreement with gravitational lensing data without dark matter. The screening of the gravitational constant ${\tilde G}$ can also be operative in MOG cosmology for the calculation of the density perturbations~\cite{Moffat2}. The monopole and dipole gravitational wave modes associated with the scalar and vector fields in MOG are screened away, leading to agreement with binary pulsar timing.

\section{MOG Field Equations}

The MOG theory has a fully covariant action composed of scalar, vector and tensor fields~\cite{Moffat1}:
\begin{equation}
\label{action1}
S=S_G+S_\phi+S_S+S_{\rm EM}+S_M.
\end{equation}
The components of the action are the Einstein gravity action ($c=1$ and the metric signature is: $(+1,-1,-1,-1)$):
\begin{equation}
S_G=\frac{1}{16\pi}\int d^4x\sqrt{-g}\Big[\frac{1}{G}(R+2\Lambda)\Big],
\end{equation}
and the massive vector field $\phi_\mu$ action:
\begin{eqnarray}
S_\phi= - \omega\int d^4x\sqrt{-g}\Big[\frac{1}{4}B^{\mu\nu}B_{\mu\nu}-V(\phi_\mu)\Big],
\end{eqnarray}
where $B_{\mu\nu}=\partial_\mu\phi_\nu-\partial_\nu\phi_\mu$ and $V(\phi_\mu)$ denotes a potential for $\phi_\mu$.  The action for the scalar fields $G$ and $\mu$ is
\begin{eqnarray}
\label{Saction}
S_S=\int d^4x\sqrt{-g}\Big[\frac{1}{G^3}\Big(\frac{1}{2}g^{\mu\nu}\nabla_\mu G\nabla_\nu G-V(G)\Big)
+\frac{1}{\mu^2 G}\Big(\frac{1}{2}g^{\mu\nu}\nabla_\mu\mu\nabla_\nu\mu - V(\mu)\Big)\Big].
\label{scalar}
\end{eqnarray}
Here, $\nabla_\mu$ denotes the covariant derivative with respect to the metric $g_{\mu\nu}$, $\omega$ is a dimensionless coupling constant and $V(G)$ and $V(\mu)$ denote potentials for the fields
$G$ and $\mu$, respectively.  In particular, we have
\begin{equation}
\label{Vphi}
V(\phi_\mu)=-\frac{1}{2}\mu_\phi^2g^{\mu\nu}\phi_\mu\phi_\nu + W(\phi_\mu),
\end{equation}
where $W(\phi_\mu)$ is the $\phi_\mu$ self-interaction.

The energy-momentum tensor is defined as
\begin{equation}
T_{\mu\nu}=T_{M\mu\nu}+T_{\phi\mu\nu}+T_{S\mu\nu}+T_{EM},
\end{equation}
where
\begin{equation}
T_{X\mu\nu}=-\frac{2}{\sqrt{-g}}\frac{\delta S_X}{\delta g^{\mu\nu}},\quad (X=[M,\phi,S,EM]).
\end{equation}
The electromagnetic (EM) action $S_{\rm EM}$ is given by
\begin{equation}
S_{\rm EM}=-\frac{1}{4}\int d^4x\sqrt{-g}F^{\mu\nu}F_{\mu\nu},
\end{equation}
where $F_{\mu\nu}$ is the EM field and the EM energy-momentum tensor is
\begin{equation}
\label{EMtensor}
T_{{EM}\mu\nu}={F_\mu}^\alpha F_{\nu\alpha}-\frac{1}{4}g_{\mu\nu}F^{\alpha\beta}F_{\alpha\beta}.
\end{equation}
Moreover, the energy-momentum tensor for the $\phi_\mu$ field is
\begin{equation}
\label{Tphi}
T_{\phi\mu\nu}=\omega\biggl({B_\mu}^\alpha B_{\nu\alpha}-\frac{1}{4}g_{\mu\nu}B^{\alpha\beta}B_{\alpha\beta}+g_{\mu\nu}V(\phi)-2\frac{\partial V(\phi)}{\partial g^{\mu\nu}}\biggr).
\end{equation}

A variation of the action with respect to $g^{\mu\nu}$ yields the field equations:
\begin{equation}
G_{\mu\nu}-g_{\mu\nu}\Lambda+Q_{\mu\nu}=8\pi GT_{\mu\nu},
\end{equation}
where
\begin{equation}
Q_{\mu\nu}=G\biggl(g^{\alpha\beta}\Theta\nabla_\alpha\nabla_\beta g_{\mu\nu}-\nabla_\mu\nabla_\nu\Theta\biggr),
\end{equation}
and where $\Theta(x)=G^{-1}(x)$.

\section{Conformal Metric Coupling with Screening}

We are interested in screening models of modified gravity with screening features in dense environments due to the non-linearity of the interaction potentials and their coupling strengths. Examples of such models are chameleons~\cite{Khoury}, dilatons~\cite{Brax1,Brax2,Brax3}, scalar field pulsar timing~\cite{Brax4} and symmetrons~\cite{Hinterbichler}. The action for the scalar field $G$ in Eq.(\ref{Saction}) can be written as
\begin{equation}
S_\chi=\int d^4x\sqrt{-g}\biggl(\frac{1}{2}g^{\mu\nu}\nabla_\mu\chi\nabla_\nu\chi - V(\chi)\biggr),
\end{equation}
where $\chi$ is related to $G$ in (\ref{Saction}) by $1/G=\chi^2/2$. The conformal metric in the Jordan frame ${\tilde g}_{\mu\nu}$ and its determinant ${\tilde g}$ are connected to $g_{\mu\nu}$ and $g$ in the Einstein frame by the conformal transformation:
\begin{equation}
{\tilde g}_{\mu\nu}=A^2(\chi)g_{\mu\nu},\quad {\tilde g}={\rm Det}{\tilde g}_{\mu\nu}=A^8(\chi)g,
\end{equation}
where $A^2(\chi)$ is the conformal factor.  The matter fields $\psi^i$ in the matter action, $S_M=S_M[\psi^i,{\tilde g}_{\mu\nu}]$, are coupled to the conformal metric
${\tilde g}_{\mu\nu}$. The total energy-momentum tensor satisfies the conservation law~\cite{Brax1,Brax2}:
\begin{equation}
\nabla_\mu T^{\mu\nu}=\frac{d\ln A(\chi)}{d\chi}(g^{\mu\nu}T-T^{\mu\nu})\nabla_\mu\chi,
\end{equation}
where $T=g^{\mu\nu}T_{\mu\nu}$.

The equation of motion for the field $\chi$ coupled to the matter energy-momentum tensor $T_{M\mu\nu}$ is given in the Einstein frame by
\begin{equation}
\label{chieqom}
\Box\chi=-h(\chi)T_M+\frac{\partial V(\chi)}{\partial\chi},
\end{equation}
where $T_M=g^{\mu\nu}T_{M\mu\nu}=\rho_M-3p_M$, and where $\rho_M$ and $p_M$ denote the density and pressure of matter, respectively.
Eq.(\ref{chieqom}) corresponds to to ``bare'' scalar field $\chi$ potential replaced by a new effective potential:
\begin{equation}
V_{\rm eff}(\chi)=V(\chi)-(A(\chi)-1)T_M.
\end{equation}
Here, $V_{\rm eff}(\chi)$ can have a global minimum in the background geometry dominated by dust matter for which $T_M=\rho_M$, $p_M=0$
and $\chi_{\rm min}=\chi_{\rm min}(\rho_M)$. The mass of the scalar field $\chi$ at the minimum, $\chi_{\rm min}$, is defined by
\begin{equation}
m_\chi^2=\frac{d^2V_{\rm eff}(\chi)}{d\chi^2}\Big|_{\chi=\chi_{\rm min}} > 0.
\end{equation}
We have
\begin{equation}
\label{Vchi}
V(\chi)=-\frac{1}{2}\mu_\chi^2\chi^2 + W(\chi),
\end{equation}
where $W(\chi)$ is a self-interaction potential.

The metric for the dust fluid in the weak field limit and in the Newtonian gauge has the form:
\begin{equation}
ds^2=(1+2\Phi)dt^2-(1-2\Phi)dx^2,
\end{equation}
where $\Phi$ is the modified (MOG) gravitational potential:
\begin{equation}
\label{MOGpot}
\Phi(\vec x)=-G_N(1+\alpha)\int d^3x'\frac{\rho({\vec x}')}{|{\vec x}-{\vec x}'|}+\alpha G_N\int d^3x'\frac{\rho({\vec x}')}{|{\vec x}-{\vec x}'|}\exp(-\mu_\phi|{\vec x}-{\vec x}'|).
\end{equation}

Consider a point mass $M$ embedded in a homogeneous background density as the source of gravity. The motion of a non-relativistic massive particle in the modified gravitational potential (\ref{MOGpot}) is screened when $\mu_\phi |{\vec x}-{\vec x}'|\ll 1$. The motion of an ultra-relativistic massless photon is not screened by the potential (\ref{MOGpot}), for the effect of the repulsive vector field $\phi_\mu$ is insignificant for a photon path~\cite{MoffatTothPhoton}. To resolve this problem, we couple the conformal metric ${\tilde g}_{\mu\nu}$ to the EM field $F_{\mu\nu}$ and the EM energy-momentum tensor (\ref{EMtensor}). Now, for the ultra-relativistic massless photon, the strength of the gravitational coupling is screened by the effective gravitational constant ${\tilde G}$ due to the mass $m_\chi(\rho)$ screening mechanism.

We have two operative screening mechanisms in MOG that complement one another. One is due to the repulsive Yukawa force mediated by the vector field $\phi_\mu$, while the second is mediated by the attractive scalar field $\chi$ (or $G$) Yukawa force. The screening due to the density dependent $\chi$ field and the $\phi_\mu$ vector field can screen the local gravitational field in the vicinity of the Sun and earth, guaranteeing the correct value for the deflection of light by the Sun (see next Section), and laboratory equivalence experiments on earth. It can also screen the gravitational radiation emitted indirectly by the binary pulsar PSR 1913-16, thereby screening away possible monopole and dipole radiation modes in the wave zone.

\section{Geodesic Equation for the Photon}

The geodesic equation for the photon in the Jordan frame is given by
\begin{equation}
\label{geodesic}
{\tilde k}^\mu{\tilde\nabla}_\mu{\tilde k}^\nu=0,
\end{equation}
where ${\tilde k}^\mu={\tilde g}^{\mu\nu} k_\nu$, ${\tilde k}^2={\tilde g}_{\mu\nu}{\tilde k}^\mu{\tilde k}^\nu=0$ and ${\tilde\nabla}_\mu$ denotes the covariant derivative with respect to the conformal metric ${\tilde g}_{\mu\nu}$. From (\ref{geodesic}) we have
\begin{equation}
\label{Geodesiceq}
\frac{d{\tilde k}^\mu}{d\lambda}+{{\tilde\Gamma}^\nu}_{\alpha\beta}{\tilde k}^\alpha{\tilde k}^\beta=0,
\end{equation}
where $\lambda$ is an affine parameter along the photon path and $d\tau^2=0$ for the proper time $\tau$. We have
\begin{equation}
{\tilde\Gamma}^\alpha_{\delta\gamma}=\frac{1}{2}{\tilde g}^{\alpha\beta}(\partial_\gamma {\tilde g}_{\delta\beta}+\partial_\beta{\tilde g}_{\delta\gamma}-\partial_\delta{\tilde g}_{\beta\gamma}),
\end{equation}
and we have
\begin{equation}
\partial_\mu{\tilde g}_{\mu\nu}=\partial_\mu[A^2(\chi)g_{\mu\nu}].
\end{equation}

The conformal coupling of the metric to $T_{\rm EM\mu\nu}$ can be interpreted in MOG as a screening of the gravitational coupling constant $G$ in the Einstein frame for $\phi_\mu\sim 0$:
\begin{equation}
\label{Gchi}
{\tilde G}=C(\chi(\rho))G.
\end{equation}
We can express this in the form~\cite{Moffat1,MoffatToth}:
\begin{equation}
\label{Galpha}
{\tilde G}=C(\alpha(\rho))G_N(1+\alpha),
\end{equation}
where
\begin{equation}
\alpha=\frac{(G_\infty-G_N)}{G_N}.
\end{equation}

The scalar field $\chi$ (or $G$ field) particle is a chameleon-like particle~\cite{Khoury} with mass $m_\chi\propto\rho_M^n$ where $n\sim 1$. The density $\rho_M$ is determined by the trace of the energy-momentum matter tensor $T_M$.

For a point mass $M$ the effective gravitational potential experienced by the photon is given by
\begin{equation}
\label{Phieq}
\Phi_{\rm eff}= \frac{{\tilde G}(\chi(\rho))M}{r}=\frac{G_N M}{r}[1+2h^2(\chi)\exp(-\mu_\chi r)].
\end{equation}
The second term in the brackets is a Yukawa-type potential (modified-force potential) obtained from the $\chi$ potential solution for a non-zero $\chi$ field mass $m_\chi$.
If $\mu_\chi r\gg 1$ or $h(\chi)\ll 1$ the local constraints of the modification of Newtonian and GR gravity are strongly suppressed. On the other hand,
if $\mu_\chi r\sim {\cal O}(1)$ or $h(\chi)\sim {\cal O}(1)$, then the deviation of the photon from its Newtonian and GR path can be significant. For an extended distribution of matter (\ref{Phieq}) becomes
\begin{equation}
\Phi_{\rm eff}=G_N\int d^3x'\frac{\rho({\vec x}')}{|{\vec x}-{\vec x}'|}[1+2h^2(\chi)\exp(-\mu_\chi|{\vec x}-{\vec x}'|)].
\end{equation}

\section{Deflection of Light in MOG}

We shall take the point particle Schwarzschild solution as the solution to the geodesic equation, valid in MOG for sufficiently large $r$~\cite{MoffatToth}:
\begin{equation}
\label{Einsteinmetric}
d\tau^2=\biggl(1-\frac{2{\tilde G}M}{r}\biggr)dt^2-\frac{dr^2}{1-\frac{2{\tilde G}M}{r}}-r^2d^2\Omega,
\end{equation}
where $d^2\Omega=d\theta^2+\sin^2\theta d\phi^2$. We have chosen the Einstein frame metric $g_{\mu\nu}$ with ${\tilde G}$ given in (\ref{Gchi}) and (\ref{Phieq}) as the effective gravitational coupling constant. We will also employ the geodesic equation in the Einstein frame:
\begin{equation}
\label{EinsteinGeodesiceq}
\frac{dk^\mu}{d\lambda}+{\Gamma^\nu}_{\alpha\beta}k^\alpha k^\beta=0.
\end{equation}
The metric (\ref{Einsteinmetric}) is approximately valid for ${\tilde G}\sim {\rm constant}$. In the solar system ${\tilde G}= G_N$, while the screening ceases to be operative in galaxies and galactic clusters and ${\tilde G}\sim G_N(1+\alpha)$.

We choose $\theta=\pi/2$ and assume initially that the photon mass $m_\gamma\neq 0$.  We have the constants of motion~\cite{Wheeler}:
\begin{equation}
\label{Energy}
\frac{E}{m_\gamma}=\biggl(1-\frac{2{\tilde G}M}{r}\biggr)\frac{dt}{d\tau},
\end{equation}
and
\begin{equation}
\label{Angularmomentum}
\frac{L}{m_\gamma}=r^2\frac{d\phi}{d\tau},
\end{equation}
where $E$ and $L$ denote the energy and the angular momentum, respectively.

From (\ref{Energy}) and (\ref{Angularmomentum}) we obtain:
\begin{equation}
dt^2=\frac{\biggl(E/m_\gamma\biggr)^2}{\biggl(1-2{\tilde G}M/r\biggr)^2}d\tau^2,
\end{equation}
and
\begin{equation}
d\phi^2=\biggl(\frac{L}{m_\gamma}\biggr)^2\frac{1}{r^4}d\tau^2.
\end{equation}
The equation of the particle orbit is given by
\begin{equation}
\biggl(\frac{1}{r^2}\frac{dr}{d\phi}\biggr)^2=\biggl(\frac{E}{L}\biggr)^2-\biggl(1-\frac{2{\tilde G}M}{r}\biggr)\biggl[\biggl(\frac{m_\gamma}{L}\biggr)^2+\frac{1}{r^2}\biggr].
\end{equation}
We now set the photon mass to zero, $m_\gamma=0$, and we get
\begin{equation}
\biggl(\frac{1}{r^2}\frac{dr}{d\phi}\biggr)^2=\biggl(\frac{E}{L}\biggr)^2-\biggl(1-\frac{2{\tilde G}M}{r}\biggr)\frac{1}{r^2}.
\end{equation}
We now find that for $L=kb$ where $b$ is the impact parameter and $L/E=b$ for $E=k$:
\begin{equation}
\biggl(\frac{1}{r^2}\frac{dr}{d\phi}\biggr)^2=\biggl(1-\frac{2{\tilde G}M}{R}\biggr)\frac{1}{R^2}-\biggl(1-\frac{2{\tilde G}M}{r}\biggr)\frac{1}{r^2},
\end{equation}
where $R$ is the distance of closest approach of the light ray. We obtain the equation for the light ray trajectory:
\begin{equation}
d\phi=d\beta\biggl[1-\frac{2{\tilde G}M}{R}\biggl(\cos\beta+\frac{1}{1+\cos\beta}\biggr)\biggr]^{-1/2},
\end{equation}
where $\cos\beta=R/r$.

The total variation of the azimuth $\phi$ along the path of the photon is given by
\begin{equation}
\phi=2\int_0^{1/2}d\beta\biggl[1+\frac{{\tilde G}M}{R}\biggl(\cos\beta+\frac{1}{1+\cos\beta}\biggr)\biggr]=\pi+\frac{4{\tilde G}M}{R}.
\end{equation}
This gives the angle of deflection $\Delta\phi$ with respect to the straight line:
\begin{equation}
\Delta\phi=\frac{4{\tilde G}M}{R}.
\end{equation}

We now use (\ref{Galpha}) to give
\begin{equation}
\Delta\phi=\frac{4{\tilde G}M}{R}=\frac{4C(\alpha(\rho))G_N(1+\alpha)M}{R}.
\end{equation}
For large scale objects such as galaxies and galactic clusters, $C(\alpha(\rho))\sim 1$, while for dense, compact objects like the Sun:
$C(\alpha(\rho))\sim 1/(1+\alpha)$, yielding for the Sun's deflection of light:
\begin{equation}
\Delta\phi=\frac{4G_NM_{\odot}}{R},
\end{equation}
in agreement with the GR prediction, $\Delta\phi=1.75^{\prime\prime}$.

\section{Lensing}

In situations where the modified gravitational theory is approximated by weak gravity, the deflection of light due to a spatially extended mass can be written as a sum over point masses. In the continuum limit, we have~\cite{Bartlemann}:
\begin{equation}
\vec\alpha(\vec \xi)=4{\tilde G}\int d^2{\vec x}'\int dz\rho({\vec x}',z)\frac{({\vec \xi}-{\vec x}')}{|{\vec \xi}-{\vec x}'|^2},
\end{equation}
where $z$ denotes the coordinate along the line-of-sight, and ${\vec \xi}-{\vec x}'={\vec b}$ is the vector impact parameter.

In the case when the distances between the source, lens and observer are much larger than the size of the lens, i.e., the ``thin lens'' limit the projected mass density is
\begin{equation}
\Sigma({\vec \xi})=\int dz\rho({\vec \xi},z),
\end{equation}
and the deflection angle is given by
\begin{equation}
\vec\alpha({\vec\xi})=4{\tilde G}\int d^2{\vec x}'\frac{({\vec\xi}-{\vec x}')\Sigma({\vec x}')}{|{\vec\xi}-{\vec x}'|^2}.
\end{equation}

The $\kappa$-convergence is defined by
\begin{equation}
\kappa({\vec\theta})=\frac{\Sigma(D_l{\vec \theta})}{\Sigma_{\rm crit}},
\end{equation}
where the critical surface density is
\begin{equation}
\Sigma_{\rm crit}=\frac{D_s}{4\pi{\tilde G}D_{ls}D_l}.
\end{equation}
Here, $D_{ls}$ is the distance from the lens to the source, $D_s$ is the distance from the observer to the source, while $D_l$ is the distance from the observer to the lens.
The deflection angle can now be expressed as:
\begin{equation}
\vec\alpha(\vec\theta)=\frac{1}{\pi}\int d^2\theta'\frac{(\vec\theta-\vec\theta')\kappa(\vec\theta')}{|\vec\theta-\vec\theta'|^2}.
\end{equation}
The $\kappa$-convergence can be written as:
\begin{equation}
\kappa(\vec\theta)=\frac{1}{2}\nabla^2\psi(\vec\theta),
\end{equation}
where $\psi(\vec\theta)$ is the deflection potential.:
\begin{equation}
\psi(\vec\theta)=\frac{2D_{ls}}{D_lD_s}\int dz\Phi(D_l\vec\theta,z),
\end{equation}
and where $\Phi$ is the MOG potential (\ref{MOGpot}).\\

\includegraphics[scale=1.0]{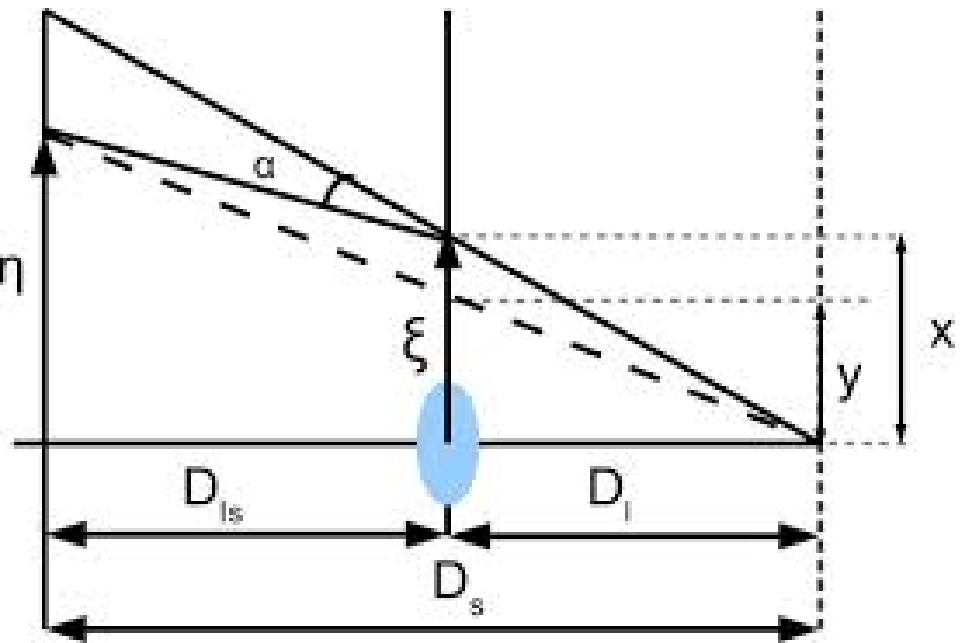}\\

Fig.1. Distances and angles involved in a gravitational lens system.

\section{Conclusions}

The scalar field $G(x)$ as well as the vector field $\phi_\mu(x)$ play a fundamental role in modified gravity (MOG) in determining gravitational interactions in early universe cosmology~\cite{Moffat2} and the present day dynamics of galaxies and clusters of galaxies~\cite{MoffatRahvar1,MoffatRahvar2}. The phion particle mass $m_\phi$ associated with Proca vector field $\phi_\mu$ evolves from a neutral cold dark matter, pressureless  particle in the early universe to an ultra-light hidden and weakly coupled photon in the present universe with a mass $m_\phi=2.6\times 10^{-28}$ eV. The potentially undetectable feature of the ultra-light hidden photon (phion) can explain the lack of success in detecting dark matter candidates such as WIMPs and axions in the universe today~\cite{Wimp,Xenon1,Lux,Xenon2}. The STVG theory (MOG) has successfully explained astrophysical data from the distance scale of the solar system to the scale of galaxies and galactic clusters without exotic dark matter. However, the long range forces associated with the scalar field $G$ and the vector field $\phi_\mu$ can lead to violations of local gravitational bounds for the solar system and the binary pulsars. To avoid this problem, two screening mechanism mediated by the $G$ and $\phi_\mu$ fields are designed to dynamically screen the forces associated with these fields in dense or high-curvature environments.

We have demonstrated how the nonlinearities of the potential couplings of the scalar degree of freedom prevent the scalar force from propagating freely in the presence of dense astrophysical objects such as the Sun, earth and neutron stars. A conformal coupling of the metric ${\tilde g}_{\mu\nu}$ in the Jordan frame to the EM field $F_{\mu\nu}$, and the screening of the gravitational coupling strength ${\tilde G}$ in the Einstein frame metric $g_{\mu\nu}$, screens the paths of photons, yielding agreement with the GR deflection of light rays grazing the limb of the Sun, and agreement with laboratory tests of the equivalence principle on earth. Moreover, the screening mechanisms screen away monopole and dipole gravitational radiation modes from sources such as binary pulsars and colliding black holes. On the other hand, the coupling strength ${\tilde G}$ enhances the strength of gravity for galaxy and cluster lensing, leading to agreement with observations, and, in particular, to agreement with the bullet cluster data~\cite{BrownsteinMoffat}.

\section*{Acknowledgments}

I thank Martin Green, Kurt Hinterbichler and Viktor Toth for helpful discussions. This research was generously supported by the John Templeton Foundation. Research at the Perimeter Institute for Theoretical Physics is supported by the Government of Canada through industry Canada and by the Province of Ontario through the Ministry of Research and Innovation (MRI).

\end{document}